# Mission Architecture to Characterize Habitability of Venus Cloud Layers via an Aerial Platform

**Rachana Agrawal** [1,*], **Weston P. Buchanan** [1], **Archit Arora** [1], **Athul P. Girija** [1], **Maxim de Jong** [2], **Sara Seager** [3], **Janusz J. Petkowski** [3], **Sarag J. Saikia** [4], **Christopher E. Carr** [5], **David H. Grinspoon** [6], **James M. Longuski** [1] **and the Venus Life Finder Mission Team**

1   School of Aeronautics and Astronautics, Purdue University, West Lafayette, IN 47907, USA;
    buchanaw@purdue.edu (W.B.); arora31@purdue.edu (A.A.); apradee@purdue.edu (A.P.G.);
    longuski@purdue.edu (J.M.L.)
2   Thin Red Line Aerospace Ltd., Chilliwack, BC V2R 5M3, Canada;
    maxim@thin-red-line.com
3   Dept. of Earth, Atmospheric, and Planetary Sciences, Massachusetts Institute of Technology,
    Cambridge, MA 02139, USA; seager@mit.edu (S.S.); jjpetkow@mit.edu (J.J.P.)
4   Spacefaring Technologies Pvt. Ltd., Bengaluru, India; saragjs@gmail.com
5   School of Aerospace Engineering and School of Earth and Atmospheric Sciences, Georgia Institute of Tech-
    nology, Atlanta, GA 30332, USA; cecarr@gatech.edu
6   Planetary Science Institute, 1700 East Fort Lowell, Suite 106, Tucson, AZ 85719-2395, USA;
    grinspoon@psi.edu
*   Correspondence: rachna.agrawal.04@gmail.com

**Abstract:** Venus is known for its extreme surface temperature and its sulfuric acid clouds. But the cloud layers on Venus have similar temperature and pressure conditions to those on the surface of Earth and are conjectured to be a possible habitat for microscopic life forms. We propose a mission concept to explore the clouds of Venus for up to 30 days to evaluate habitability and search for signs of life. The baseline mission targets a 2026 launch opportunity. A super-pressure variable float altitude balloon aerobot cycles between the altitudes of 48 and 60 km, i.e., primarily traversing the lower, middle, and part of the upper cloud layers. The instrument suite is carried by a gondola design derived from the Pioneer Venus Large Probe pressure vessel. The aerobot transmits data via an orbiter relay combined with a direct-to-Earth link. The orbiter is captured into a 6-h retrograde orbit with a low, roughly 170-degree, inclination. The total mass of the orbiter and entry probe is estimated to be 640 kg. An alternate concept for a constant float altitude balloon is also discussed as a lower complexity option compared to the variable float altitude version. The proposed mission would complement other planned missions and could help elucidate the limits of habitability and the role of unknown chemistry or possibly life itself in the Venus atmosphere.

## 1. Introduction

The search for signs of life in our solar system has been made possible by advanced robotic exploration. The discovery of subsurface oceans on Europa, salty plumes on Enceladus, methane lakes on Titan, and more recently, the potential presence of phosphine in the Venusian clouds [1,2], has increased the expectation that these planets and moons might harbor extant life.

The cloud layers of Venus, which extend from 48 km to 70 km, present temperature and pressure conditions similar to those on the surface of the Earth—contrasting markedly with the surface environment of 450 °C and 90 bar pressure. For decades, the clouds of Venus have been thought to be potentially habitable and even possibly harboring life [3]. The topic of life on Venus has regained popularity in light of both longstanding and recent discoveries, which include the unknown composition of Mode 3 cloud particles [4]; the "mysterious UV absorber" in the upper clouds [5]; the presence of tens of ppm of $O_2$ [6,7]; the potential presence of $PH_3$ [1]; the $SO_2$ and $H_2O$ vertical abundance profiles [8]; and the possible presence of $NH_3$ [9]. These anomalies, some of which have observably lingered for decades, might be tied to habitability and life's activities or be indicative of unknown chemistry which, in its own right, is worth exploring.

These findings warrant a dedicated exploration of the Venus cloud decks to assess habitability, look for signs of life or even life itself, and lay the groundwork for an eventual atmospheric sample return mission.

In this paper, we present in detail one version of the architecture of the Venus Life Finder (VLF) "medium" mission, which would build upon a targeted probe mission in development (further details in [10]) that seeks evidence of a carbon cycle in the Venus atmosphere and will spend only a few minutes within the temperate cloud zone. The VLF medium mission is a balloon mission that enables more sophisticated science instruments to operate over a longer duration than an atmospheric probe, providing spatial and temporal variability in sampling.

The only precedent of a balloon mission to Venus, or any planetary body, is the VeGa mission sent by the Soviet Union in 1985 [11]. The VeGa mission consisted of two identical probes to Venus, each carrying a balloon platform and a lander. The balloons successfully transmitted the science data back to Earth. The zero-pressure balloon floated at 54 km altitude, transmitting data for about 48 h, and traversed about 11,000 km around the planet, close to the equator. The inflated balloon was about 3 m in diameter, with a total balloon and gondola mass of 23 kg. The gondola was a 6 kg system, with atmospheric structure instruments including temperature, pressure, and wind velocity sensors. The gondola also had a backscattering nephelometer [12]. The overall success of the mission set a precedent for using balloon technology on Venus (and elsewhere in the solar system) and provided a tangible baseline for future missions.

Several balloon mission concepts have been designed since the VeGa mission to explore the clouds of Venus. Some of these include the Venus Climate Mission [13], European Venus Explorer [14], Balloon Experiment at Venus [15], Buoyant Venus Station [16], and the 2020 Venus Flagship Mission [17]. We summarize them briefly for context as follows. The Buoyant Venus Station is one of the early balloon concepts for Venus exploration using a fixed-float altitude super-pressure balloon with a mission life ranging from one week to several months. The authors suggest multiple possible types of missions including flyby, orbiter, and swing-by. The Venus Climate Mission, a planetary decadal study by NASA, deploys a balloon with a float altitude of 55.5 km. The Venus Climate Mission balloon carries a mini probe and two drop sondes that are deployed at different times over a nominal operational time of 21 days [13]. The balloon returns data via an orbiter in Venus orbit that acts as a communication relay. The European Venus Explorer, a mission proposal to ESA, consists of a balloon with a float altitude of 55 km and an operation period of 10 days [14]. The European Venus Explorer balloon returns data via a combination of a direct-to-Earth link and a link with the flyby carrier spacecraft. NASA JPL's Balloon Experiment at Venus concept consists of a variable altitude balloon that uses a reversible fluid altitude control technology [15]. The balloon cycles between 40 and 60 km every six hours. The balloon uses a direct-to-Earth strategy to return science data. The data return strategy in each of the above concepts correlates to the size of the science payload and science data.

In this paper, we present the mission architecture for one version of the VLF medium mission. First, we introduce the science requirements, which can be achieved with a balloon platform carrying a gondola with science instruments, with a mission life of two weeks and a possible extension to 30 days. We present two balloon platform concepts: (1) Constant Float Altitude (CFA) and (2) Variable Float Altitude (VFA). The CFA balloon has a fixed-float altitude of 52 km whereas the VFA balloon cycles between the altitudes of 48 km and 60 km. In this paper, we describe the VFA concept in detail and provide a comparison with the CFA in Section 8.

In this paper, we focus on the VFA balloon concept. The VFA balloon is a simple super-pressure balloon that, in the context of the herein considered ConOps, varies altitude between 48 and 60 km. To change altitude, the balloon uses an innovative combination of (a) conventional ballast drop and controlled gas venting, and (b) strategic use of heating and cooling associated with the Venus diurnal cycles. The VFA concept all but eliminates the complexity of a Variable Altitude Balloon (VAB) that incorporates active mechanical or pump-based density control systems while still facilitating a reasonable number of altitude cycles to accomplish the intended science objectives. The balloon returns data using a combination of an orbiter communication link and direct-to-Earth transmission.

## 2. Science Goals and Requirements

The science focus of the mission is to investigate the habitability of cloud particles. The science goals are:

1.  Measure habitability indicators of the cloud deck:
    -   Determine the amount of water vapor in the cloud layers
    -   Determine the acidity of the cloud droplets
    -   Detect and identify metals in the cloud particles
    -   Measure temperature, pressure, and windspeed
2.  Search for evidence of life in the Venusian clouds:
    -   Detect reduced and anomalous molecules as a sign of disequilibrium chemistry and as biosignature gases
    -   Detect presence of organic material within cloud-layer particles
    -   Identify organic material within cloud-layer particles
3.  Characterize cloud particles in preparation for a sample return mission:
    -   Determine if cloud particles are liquid or solid or both
    -   Determine the level of homogeneity of the cloud particles

The altitude range of 48 to 60 km is chosen for science operations as it spans all three cloud layer regions: the lower, middle, and part of the upper cloud layer. The lower cloud layers, which are situated in the 48 to 52 km altitude range are known to contain the majority of the so-called "Mode 3" particles of unknown composition [4]. The data from the Pioneer Venus probe suggest that some cloud particles are not spherical and as such cannot be composed solely of pure liquid sulfuric acid [4] leading to the possibility of unknown chemistry, including life [9,18]. The middle cloud layer, dominated by spherical, liquid droplets of sulfuric acid stretches from 51 to 56 km. The upper cloud layer, above 56 km, contains the so-called "unknown UV absorber" [5]. The UV absorber has properties that are suggested to be consistent with biological activity by some researchers [19], while others have suggested potential abiotic matches, such as $S_2O_2$ [20].

*Science Instruments*

Our philosophy for the VLF mission is to send a suite of small, low-complexity science instruments to operate just long enough to assess the cloud-layer habitability, search for biosignatures, and to prepare for sample return. The science instrument suite is listed in Table 1, with a mass, power, and volume summary. These instruments are housed inside the balloon gondola. The power listed corresponds to an operation cycle of the respective instruments. Over the course of the mission, the average power requirement of an instrument will depend on the total operation time. At a given time, only one instrument is operated. On average, we assume that each instrument operates for roughly 20% of the mission duration to provide a baseline power requirement.

**Table 1.** Science instruments considered for the gondola.

| Instrument ** | Mass (kg) | Volume (cm³) | Average Power (W) | Current TRL * |
|---|---|---|---|---|
| Mini Tunable Laser Spectrometer (mTLS) | 4.60 | 240 | 14.0 | 6 |
| Ion-Gas Micro-Spectrometer for Aerosols (MEMS-A) | 0.34 | 400 | 1.0 | 4 |
| Auto-fluorescing Nephelometer (AFN) | 0.80 | 100 | 40.0 | 3 |
| Tartu Observatory pH sensor (TOPS) | 0.35 | 844 | 2.0 | 2 |
| Imaging Unit (IU) | 0.15 | 250 | 0.5 | 5 |
| Weather Instruments Suite (WIS) | 0.10 | 98 | 1.0 | 5 |
| **Total Gondola Science Instruments** | **6.34** | **1932** | **58.5** | |

\* Technology Readiness Level (TRL) does not account for Venus clouds as a relevant environment.
\*\* For a detailed discussion of the instruments and science behind the proposed habitability mission architecture, see the VLF mission study report [10].

There are also four mini probes, each containing a different instrument as well as a Weather Instruments Suite (WIS). The four instruments include an Auto-fluorescing Nephelometer (AFN), MEMS Aerosol Elemental Analyzer (MEMS-A), Tartu Observatory pH Sensor (TOPS), and another pH sensor concept called the Molybdenum Oxide Sensor Array (MoOSA). The total instrument mass for the mini probes is about 1.6 kg. For a detailed discussion of the instruments and science behind the proposed habitability mission architecture, see the VLF mission study report [10].

## 3. Mission Architecture Overview

The mission architecture primarily consists of three elements launched in a stacked configuration on the same vehicle. These elements are:

1. An aerial platform that consists of the balloon, gondola, mini probes, and reserve gas tank. The aerial platform cycles between 48 and 60 km altitude and can survive for more than 30 days in the cloud layers. The platform deploys mini probes at various altitudes and times.
2. An entry probe that carries the aerial platform and inflation system. The system survives a peak heat rate of ~1.5 kW/cm² and peak deceleration of 69 g and deploys the aerial platform at around 50 km altitude.
3. An orbiter that carries the entry probe to Venus and functions as a communication relay. The orbiter provides power and communications support to the entry probe during the cruise phase. The orbiter relays the data received from the gondola to the Earth ground station.

The stacked orbiter-probe spacecraft is launched from Earth in July 2026 and reaches Venus after about 120 days. We choose the 2026 launch date to meet one of the goals of the VLF mission study—to identify the earliest possible opportunity to accelerate and catalyze further exploration of the Venus atmosphere. The probe separates from the orbiter about 30 days before entering the Venus atmosphere. The orbiter inserts into a Venus orbit with a 6-h period and 1000 km periapsis altitude. Figure 1 shows a schematic of the mission architecture.

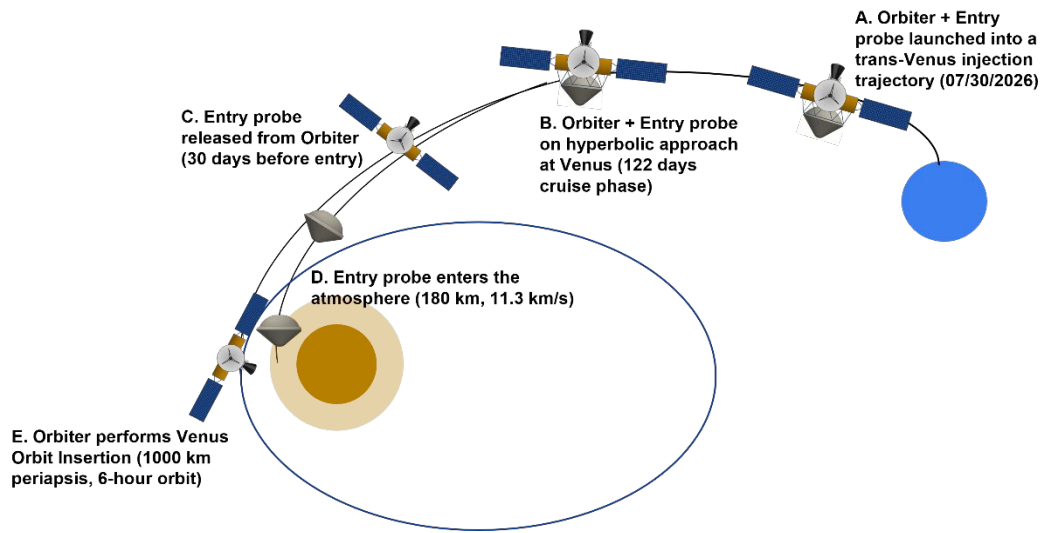

**Figure 1.** Overview of the launch and arrival operations. Launch occurs on 30 July 2026; Probe enters atmosphere and orbiter captures in orbit after 122 days.

## 4. Mission Design

The mission operations are split into four phases: (1) Launch and Cruise to Venus, (2) Venus Approach, (3) Venus Orbit Insertion and Entry, Descent, and Inflation, and (4) Nominal Balloon Operations.

### 4.1. Launch and Cruise

The baseline mission launches on 30 July 2026. The possible launch vehicles include Rocket Lab's Neutron Rocket and SpaceX's Falcon 9 with the first stage recovered via the SpaceX Automated Spaceport Drone Ship (ASDS). The trajectory is a direct Earth-to-Venus transfer with a launch $C_3$ of 7.12 km²/s² and a time of flight (TOF) of 122 days. The spacecraft arrives in the vicinity of Venus on 29 November 2026 with a $V_\infty$ of 4.92 km/s. There are alternate launch opportunities in 2026 that leverage a 200 km lunar flyby upon departure from Earth. There are two such trajectories with launch $C_3$s of 5 km²/s² and 6 km²/s², and with TOF and arrival $V_\infty$ similar to that of direct transfer. Such trajectories increase the delivered mass and can reduce the overall mission cost, but also add complexity.

### 4.2. Venus Approach

Thirty days prior to entering the Venus atmosphere, the entry probe separates from the orbiter and continues towards the targeted entry location. This separation imparts a small impulse on the orbiter, and so two days are allotted for orbit determination following entry probe separation. After the orbit determination period, the orbiter performs a propulsive deflection maneuver with a $\Delta V$ of up to 200 m/s to raise its periapsis out of the atmosphere to an altitude of about 1000 km and adjust the relative geometries of both spacecraft so that the orbiter is in line of sight with the balloon during entry and inflation. Figure 2 shows the arrival geometry of the spacecraft.

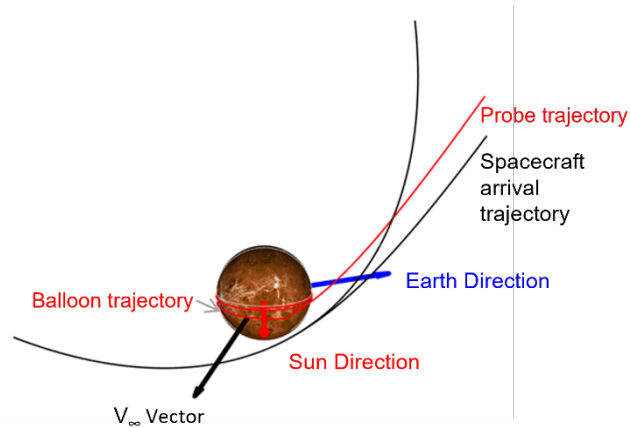

**Figure 2.** Arrival geometry of probe and orbiter. The arriving spacecraft on a trajectory to enter the atmosphere. The orbiter releases the probe 30 days before entering the atmosphere and performs a deflection maneuver to raise the altitude above the atmosphere.

The probe entry location is determined by various factors. Figure 3 shows the 2-D projection of the probe entry locations colored based on the feasibility. The entry flight path angle depends on the arrival geometry and, in turn, determines the latitude regions accessible. The potential entry region is selected such that the probe is in line of sight with Earth during entry and the balloon deployment is near the equator so that the balloon trajectory does not drift significantly along the longitude direction.

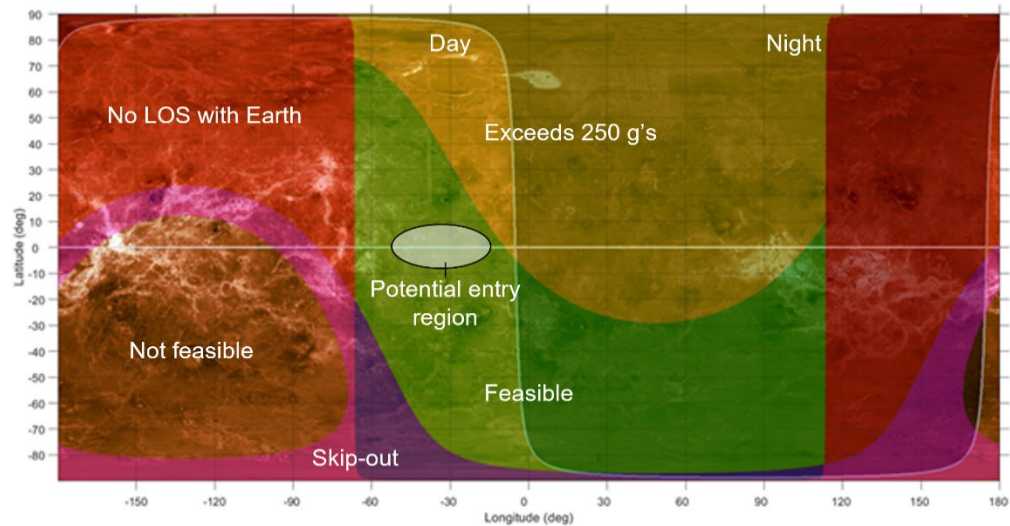

**Figure 3.** Entry location accessibility map for the 2026 trajectory. Green: Feasible entry region with a line of sight with Earth; Red: No line of sight with Earth; Yellow: Deceleration load greater than 250 Earth g; Pink/Purple: Probe skips out of the atmosphere. The probe enters near the equator within the region marked by the oval.

### 4.3. Entry, Descent, Inflation, and Orbit Insertion

The probe enters the atmosphere at a speed of 11.33 km/s at an entry flight-path angle (EFPA) of −10 degrees. The entry interface is defined at 180 km above the surface. The shallow EFPA is chosen to aid data relay to the orbiter and minimize peak deceleration as well as stagnation-point heat rate. Figure 4 shows the concept of operations (conops) for entry, descent, and inflation for the entry system. Table 2 shows the timeline of key events during entry, descent, and inflation.

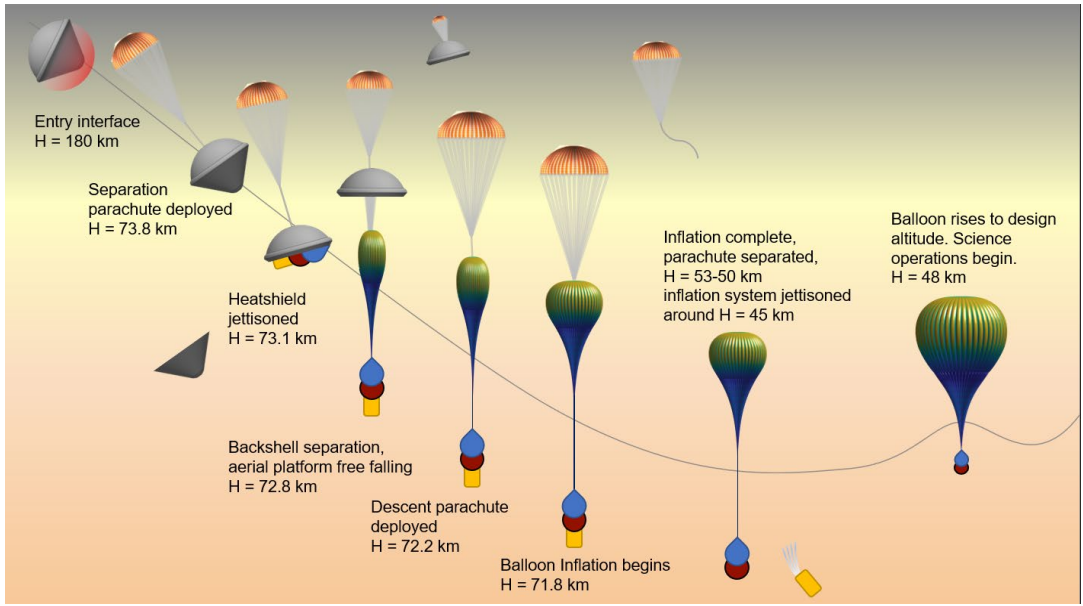

**Figure 4.** Entry, Descent, and Inflation concept of operations. The EDI conops is derived from the VeGa balloon EDI conops [21].

Following peak heating and peak deceleration, the separation parachute is deployed at around 74 km altitude. The total estimated stagnation-point heat load is 18 kJ/cm². Five seconds later, the front heat shield is jettisoned and another five seconds later the aerial platform payload is deployed from the backshell. After five seconds of free fall, the aerial platform descent parachute deploys, and the inflation system begins filling the balloon envelope at an altitude of 72 km. The inflation is complete at around 52 km and the descent parachute is released. After ten minutes, the inflation system is jettisoned, the balloon's flight control system is engaged, and the balloon begins variable altitude science operations. Figure 5 plots altitude vs. time, deceleration, and heating profiles for the VLF medium habitability mission.

**Table 2.** Timeline of the entry, descent, and inflation operations.

| Time, s | Altitude, km | Speed, km/s | Event |
|---|---|---|---|
| E + 0 | 180 | 11.33 | Entry interface |
| E + 56 | 88.5 | 9.59 | Peak heating, heat rate = 1502 W/cm² |
| E + 60 | 84.7 | 7.15 | Peak deceleration, 69 g |
| E + 100 | 73.8 | 0.381 | Separation parachute deployed |
| E + 105 | 73.1 | 0.204 | Front heatshield separation |
| E + 110 | 72.8 | 0.155 | Backshell separation, aerial platform free fall |
| E + 115 | 72.2 | 0.103 | Descent parachute deployed |
| E + 120 | 71.8 | 0.102 | Balloon inflation begins |
| E + 8 min | 53.8 | 0.028 | Balloon inflation complete, parachute cut-off |

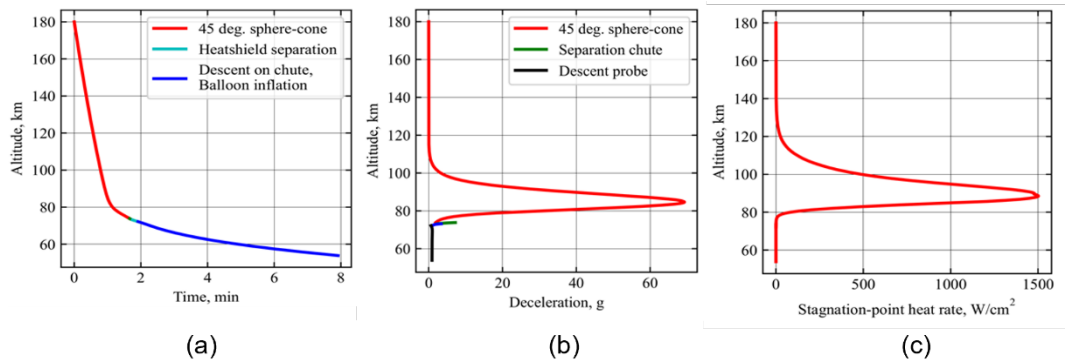

**Figure 5.** (**a**) Altitude vs time. (**b**) Deceleration profile showing peak deceleration = 69 g. (**c**) Aerodynamic heating profile showing peak heat rate = 1502 W/cm². The deceleration load and heat rates can be reduced with a shallower entry and lower ballistic coefficient of entry vehicle.

The orbiter performs its insertion maneuver to capture into an approximately 6-h orbit around Venus. The magnitude of the insertion maneuver is about 2400 m/s.

## 5. Aerial Platform Concept of Operations

The super-pressure balloon is carried by Venus's super-rotating atmosphere around the planet and circumnavigates in roughly five days, completing about six circumnavigations over 30 days. The thermal effect of the Venus diurnal cycles on the balloon's lifting gas density is a key contributor to the altitude control strategy of the aerial platform. In careful concert with diurnal effect timing, altitude change is affected by ballast release to increase altitude and venting of lifting gas to initiate a descent. The inflation system and mini probes act as ballast, with the probes enhancing science return at the same time.

The nominal conops of the aerial platform is shown in Figure 6 and are described below:

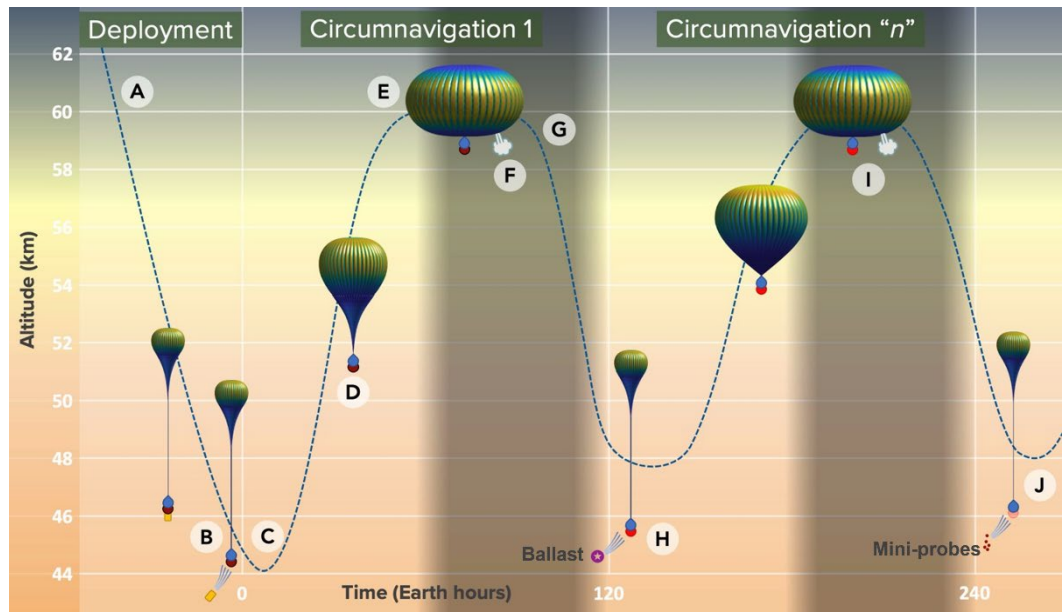

**Figure 6.** Variable Altitude Balloon (VAB) aerial platform conops. See the main text for description of steps A–J.

A.   At approximately 73 km altitude, the heat shield and backshell separate, and the aerial platform is released. The descent parachute is deployed at roughly 72 km and balloon inflation begins.

B.  Balloon inflation is complete as described above. The aerial platform continues to fall under the mass burden of the inflation system which is released only after inflation is complete. The aerial platform briefly passes below 48 km altitude, at which point the inflation system is released, followed by the jettison of the descent parachute.

C.  Relieved of inflation system mass, the fully inflated aerial platform begins to ascend.

D.  The aerial platform ascends, with the lifting gas warming in the midday Venus environment. If and where (altitude-wise) desired, ascent velocity can be restrained to near-static values through gas venting in concert with float super-pressure algorithms as described in step "F" to follow. PID (proportional–integral–derivative) type autonomous ballast and gas control are implemented to effectuate desired rates of ascent, descent, and "hold" in response to atmospheric condition variability.

E.  The maximum float altitude of 60 km is attained. The stiffness of the fully distended balloon envelope prevents the aerial platform from exceeding the target float altitude. The 60 km float altitude can be maintained for the duration needed to achieve science objectives.

F.  The balloon's maximum super-pressure is reached at 60 km float during the Venus "afternoon" and is preemptively calibrated and operationally fine-tuned through minor gas venting. This maximum super-pressure is quantified such that the subsequent nighttime cooling negates the positive buoyancy associated with the daytime super-pressure, only just to the extent that the vehicle is rendered marginally negatively buoyant, thereby setting the stage for the descent cycle.

G.  Late in the Venus night, the lifting gas reaches thermal equilibrium—cooling to its minimum temperature—leading to loss of positive buoyancy and initiation of descent. The descent continues unimpeded in a nominally stable atmosphere.

H.  PID altitude control "ramping" slows descent as the mission's 48 km minimum altitude is approached. The ramping process that slows descent and reverses the downward trajectory makes use of ballast jettison or mini probe release for the requisite return to positive buoyancy. In view of the needed reversal of trajectory, a neutrally buoyant transient float can be maintained at minimum altitude as needed for science objectives. Mini probe release is schematically pictured in Figure 6.

I.  The maximum float altitude of 60 km is regained, and balloon buoyancy is autonomously recalibrated further to steps "D" and "F" earlier. Supported by lifting gas venting on an as-needed basis, nighttime cooling is used to the maximum extent to reinitiate the descent trajectory. The maximum float altitude can also be reduced on subsequent cycles depending on science or operational needs.

J.  The process described in step "H" is repeated, schematically showing ballast jettison.

## 6. Flight Systems

The major flight systems are described in further detail in this section. To reduce the design, development, testing, and evaluation (DDT&E) cost, we use heritage or high TRL subsystems where possible. With the current progress in CubeSat technology, spacecraft subsystems are available at a low cost. Using heritage technology and design also reduces the mission risks. Table 3 summarizes the mass estimates of major flight systems. The payload capacity to Venus of Rocket Lab's Neutron Rocket is about 1500 kg and that of SpaceX's Falcon 9 is about 4000 kg, providing 80% and 400% mass margin respectively. The huge mass margin would allow for ridesharing, which could reduce the launch vehicle cost.

**Table 3.** Summary of mass of flight system elements.

| System | CBE * (kg) | Contingency (%) | MEV * (kg) |
|---|---|---|---|
| Orbiter platform | 400 | 30% | 520 |
| Entry System | 156 | 30% | 203 |
| Aerial Platform + Mini probes | 84 | 30% | 109 |
| **Total Stacked Mass** | **640** | **30%** | **832** |

* CBE is current best estimate and MEV is the maximum expected value.

### 6.1. Orbiter

The primary function of the orbiter is to act as a communication relay between the aerial platform and the Earth ground station. The orbiter is a small satellite that leverages small spacecraft technologies such as the Mars Cube One (MarCO) CubeSats that acted as a relay between the InSight Lander on Mars and the Earth ground station [22]. The orbiter includes a propulsion stage for the Venus Orbit Insertion maneuver. The sizing assumes a solid rocket motor. The mass can be improved by using a bipropellant motor. Table 4 shows the mass estimates of the orbiter subsystems. Figure 7 shows the schematic of the orbiter.

**Table 4.** Mass breakdown of the orbiter.

| Subsystem | CBE (kg) | Contingency (%) | MEV (kg) |
|---|---|---|---|
| Structures | 66 | 30% | 86 |
| Comms, Power, OBC | 55 | 30% | 72 |
| Rocket Motor and Tanks | 22 | 30% | 29 |
| **Orbiter Dry Mass** | **143** | **30%** | **186** |
| Propellant | 257 | 30% | 334 |
| **Orbiter Wet Mass** | **400** | 30% | 520 |

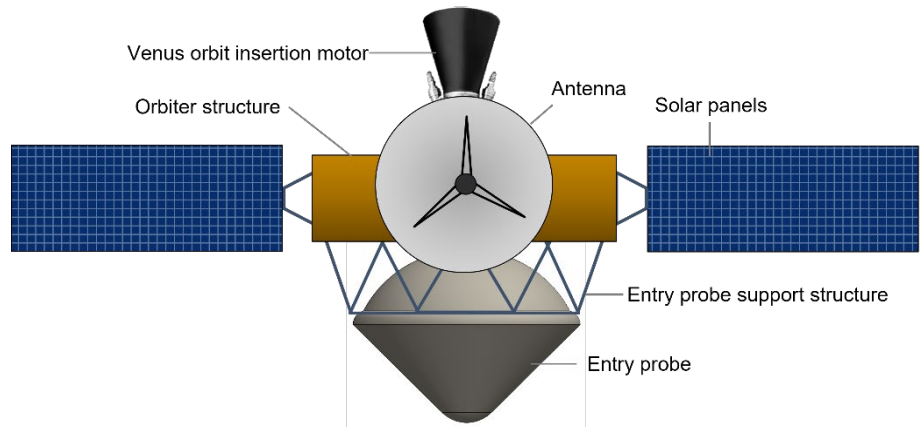

**Figure 7.** A notional schematic of the orbiter with the probe mounted.

### 6.2. Entry Probe

The VLF medium habitability mission entry probe geometry is based on the Galileo and Pioneer Venus entry probes [23]. The probe has a maximum diameter of 1.5 m and has a traditional 45-degree sphere-cone geometry. The major components of the probe are shown in Figure 8. The descent probe houses the VFA balloon, eight inflation tanks, two parachutes, a reserve gas tank, mini probes, and a gondola. There is no enclosure for the descent module. Some structural elements, however, are required to hold the different elements together inside the entry vehicle.

The entry probe uses two parachutes, both of which are based on the parachutes used on the Galileo probe [24]. The separation parachute is a 6 m diameter conical ring sail design which will allow the front heatshield and structure to be jettisoned when the parachute is released. The descent module then separates from the backshell and the separation parachute, and after a few seconds of free fall inflates the descent parachute which is a 3 m diameter parachute. The descent probe parachute is sized such that it allows the probe to spend at least 6 min before it reaches an altitude of 52 km.

**Table 5.** Mass breakdown of the entry probe.

| Subsystem | CBE (kg) | Contingency (%) | MEV (kg) |
|---|---|---|---|
| Forebody TPS (HEEET) | 50 | 30% | 65 |
| Forebody Structure | 26 | 30% | 34 |
| Backshell TPS (PICA) | 10 | 30% | 13 |
| Backshell Structure | 12 | 30% | 16 |
| Separation Parachute and Mortar | 12 | 30% | 16 |
| Separation System | 10 | 30% | 13 |
| **Aeroshell Total** | **120** | **30%** | **156** |
| Inflation System | 25 | 30% | 33 |
| Descent Parachute | 6 | 30% | 8 |
| Aerial Platform Flight System | 84 | 30% | 109 |
| Engineering System | 5 | 30% | 7 |
| **Descent Module Total** | **120** | **30%** | **156** |
| **Entry Probe (Aeroshell + Descent Module) Total** | **240** | **30%** | **312** |

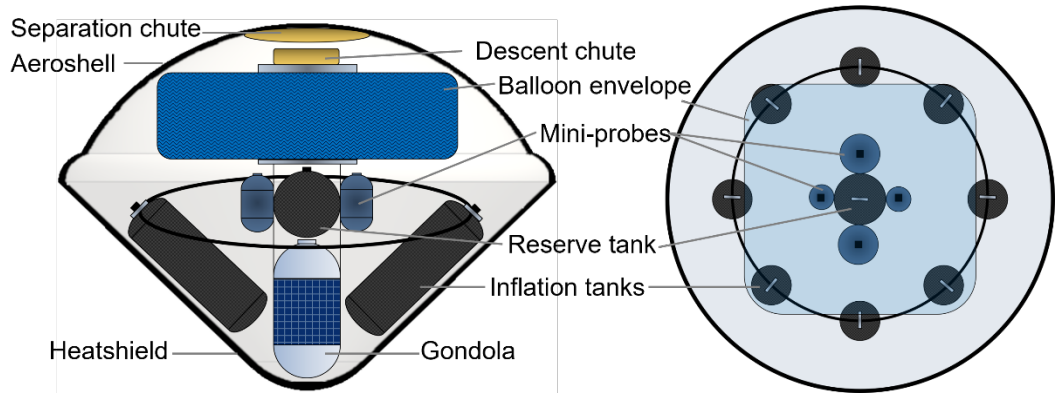

**Figure 8.** Schematic of the entry probe showing the notional accommodation of the subsystems inside the probe. (**Left**) Side view, (**Right**) Top view.

*6.3. Aerial Platform*

The helium-filled super-pressure balloon is 8 m in inflated diameter and 36 kg in mass (excluding the inflation system). The balloon envelope incorporates an innovative 5-layer metalized gas barrier laminate for long-term retention of the balloon's lifting gas, and an acid-resistant fluoropolymer exterior. The oblate spheroidal balloon architecture is a so-called "Ultra-High-Performance Vessel" (UHPV)—a highly unique, proprietary inflatable pressure vessel architecture that has been the focus of numerous NASA contracts ranging in topic from space habitats to cryogenic propellant tanks [25]. It is widely considered to be the most structurally predictable, lowest specific mass containment architecture (collapsible, metallic, composite, or otherwise), furthermore featuring unlimited scalability and structural determinacy.

**Table 6.** Mass breakdown of the aerial platform.

| Subsystem | CBE (kg) | Contingency (%) | MEV (kg) |
|---|---|---|---|
| Balloon Envelope | 37 | 30% | 48 |
| Ballast/reserve gas | 15 | 30% | 20 |
| Helium | 2 | 30% | 3 |
| Gondola | 22 | 30% | 29 |
| Mini probes | 8 | 30% | 10 |
| **Aerial Platform Total** | **84** | **30%** | **109** |

Both constant and variable altitude versions of the balloon have been successfully flight-demonstrated in Earth's atmosphere. The inflation mechanism for the Venus application is low TRL due to the conditions of deployment, which include a rapid inflation rate, descent velocity potentially greater than 10 m/s, sulfuric acid clouds, and autonomous operations. While the balloon construction and operation are proprietary to Thin Red Line Aerospace, further information can be found in [26].

*6.4. Gondola*

The gondola is a cylindrical pressure vessel based on Pioneer Venus Large Probe heritage [27]. The structure is about 20 cm in diameter and about 40 cm high and is made with titanium. The instruments are placed inside the vessel on a beryllium shelf, to provide mass-efficient high thermal conductivity and capacitance to buffer against thermal fluctuations. Figure 9 shows a notional gondola design. We choose the pressure vessel design to protect the instruments and other subsystems and electronics from the sulfuric acid clouds of Venus. The vessel reduces the technology development cost of making all the components resistant to sulfuric acid at the expense of increased structural mass. Replacing the pressure vessel with an aluminum structure saves about 1 kg of mass. With advancements in materials technology, the mass of the pressure vessel can be reduced by using lightweight materials such as beryllium or composites [28]. Since the gondola remains in relatively benign environments (below 150 °C and 6 bars), extreme environment technology development is not required.

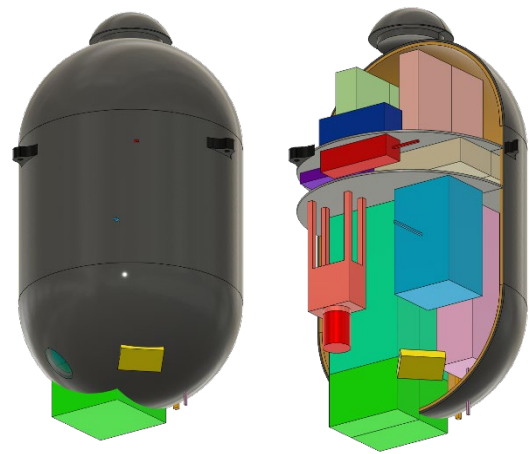

**Figure 9.** A notional schematic of the gondola pressure vessel with the instruments and subsystems placed on shelves inside the vessel.

*Thermal Control:* The gondola pressure vessel is lined with layers of Kapton on the inside. The beryllium shelf and instruments deck are separated by a Phase Change Material (PCM) to insulate the instruments further. Sodium Silicate PCM is used. The balloon cycles between 48 and 60 km, giving it time to cool down periodically. For this study, we consider the passive thermal control design sufficient for such operations.

*Power:* We use a combination of solar panels and primary and secondary batteries throughout the mission lifetime. We use a Lithium Carbon Monofluoride (LiCFx) battery pack, which has 450 Wh/kg of specific energy. The total mass of required batteries is 4 kg, and the solar panel mass is 0.3 kg. Including a power distribution system and contingencies, the total power subsystem weight is about 5 kg. The energy requirement for the mission is deduced from a baseline science and communications conops. The solar panels are body-mounted on the outside of the gondola and generate about 12 W of power on average during the day. Solar panels allow for extended operations in the cloud layer.

*Communication:* The antenna is a crossed dipole designed for 2.4 GHz S-band transmission. The antenna is mounted on top of the gondola. The balloon is opaque to radio transmission and thus about 10 degrees elevation from the zenith is blocked for communication. The dipole antenna has a wide beamwidth, which allows for transmission at elevations above 30 degrees. A breakdown of the gondola system mass is provided in Table 7.

**Table 7.** Mass breakdown of the balloon gondola.

| Subsystem | CBE (kg) | Contingency (%) | MEV (kg) |
|---|---|---|---|
| Structure | 3 | 30% | 3 |
| Science instruments | 7 | 30% | 8 |
| Battery + Power Distribution System | 5 | 30% | 7 |
| Communication | 4 | 30% | 5 |
| Thermal | 2 | 30% | 2 |
| Command & Data Handling | 3 | 30% | 4 |
| **Gondola Total** | **22** | **30%** | **29** |

Gondola Instruments Accommodation

The instruments are mounted on two shelves stacked vertically. There are three inlets that intake the atmospheric samples: one for the Mini Tunable Laser Spectrometer (mTLS), one for the AFN, and a third for the MEMS-A. There is one window looking out into the atmosphere for the camera. There are three protrusions from the gondola that do not provide visual access or intake samples. These protrusions are for the WIS, including one for the temperature and pressure sensor, one for the anemometer, and a third for the TOPS pH sensor. Glass covers seal the inlets and are ejected after deployment when desired. Figure 10 shows the various instruments and subsystems inside the gondola. The location of the inlets and TOPS protrusion depends on the flow of the atmosphere around the gondola and needs high fidelity analysis in order to facilitate the detailed design.

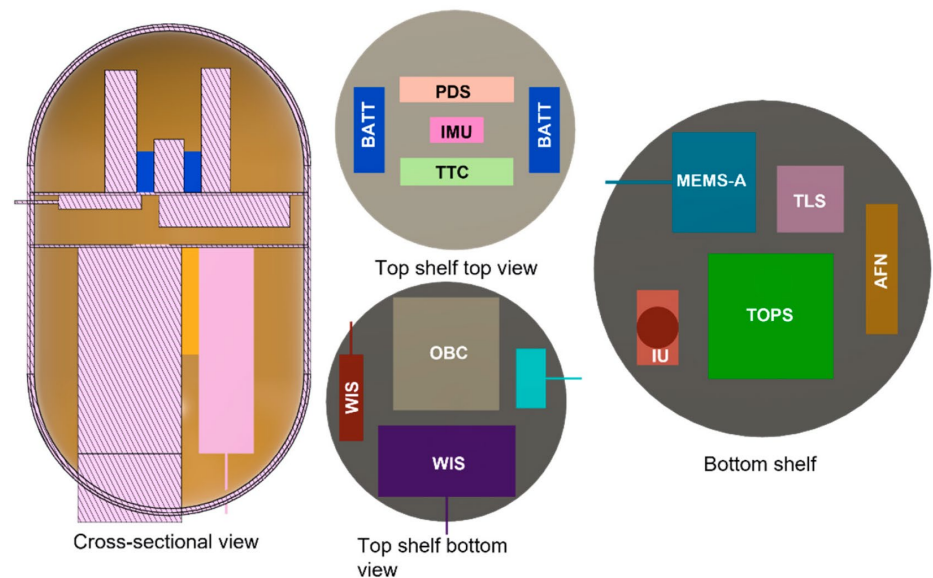

**Figure 10.** A notional schematic of the gondola with the instruments and subsystems mounted on two shelves. See Table 1 for the list of science instruments. For a detailed discussion of the instruments and science behind the proposed habitability mission architecture, see the VLF mission study report [10]. The subsystems are—BATT: Battery; TTC: Telemetry, Tracking, and Command; OBC: Onboard Computer; IMU: Inertial Measurement Unit. Instrument acronyms are defined in Table 1.

## 7. Communication Architecture

We choose a combination of direct-to-Earth and orbiter relay to return the science data. The balloon transmits at a rate of 100 bps direct-to-Earth. The orbiter relay data rate is shown in Figure 11. The total data volume transmitted over a period of 14 days is about 250 MB via the orbiter and about 25 MB via direct-to-Earth transmission. The data volume does not account for improvements from compression. With lossless compression algorithms and data packing, a compression ratio of up to 3 can be achieved. A higher compression ratio can be achieved if some loss in data is acceptable. With the development of more efficient transmission bands such as the Ka-band, the data rate can be improved.

The number of contacts and data rate to the relay depend on the orbit. Thus, the orbit is an important variable in the trade study. The orbiter is in a 6-h orbit, with low retrograde inclination and periapsis in the hemisphere of the balloon's deployment location. A retrograde orbit is chosen because the rotation of the atmosphere results in the balloon circumnavigating the planet in a retrograde direction. A low inclination allows for the orbiter to be within line of sight with the balloon for the maximum duration of the orbit. The selected orbit provides roughly 16 passes over the duration of one circumnavigation of the balloon, each of about 8 to 10 min. Figure 11 shows the elevation of the orbiter with respect to the balloon along with the range and data rate for the passes. The orbit insertion can be controlled such that the passes during the first circumnavigation are timed as desired.

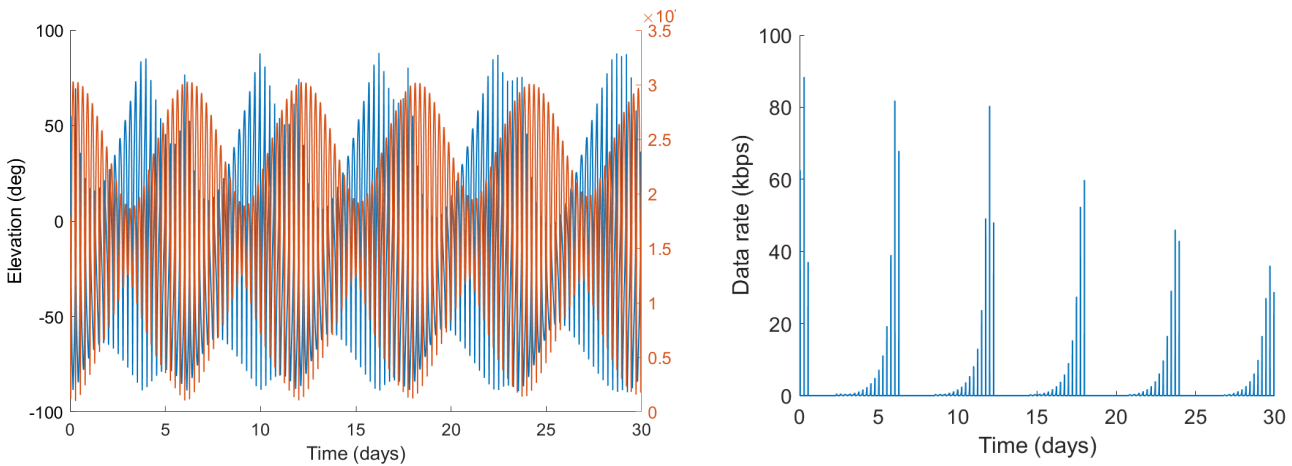

**Figure 11.** Communication link between balloon and orbiter. Left: Elevation and range of orbiter with respect to the balloon. Right: Data rate of downlink over 30 Earth days. The orbiter has about 16 contacts with the balloon over one circumnavigation, each of roughly 8 to 10 min duration. We assume that the balloon circumnavigates Venus in six days and drifts 10 degrees longitudinally over 30 days. The drift is calculated based on Venus GRAM wind models.

We also analyzed a synchronous orbit for the relay which has a period equal to the circumnavigation period of the balloon. A synchronous orbit increases the contact window with the balloon. The challenge is that the balloon's trajectory is not well-modeled and depends on the atmospheric circulation, which is not yet highly characterized. A small deviation in the balloon's trajectory can lead to the orbiter and balloon losing contact.

## 8. Constant Float Altitude Balloon Mission

The CFA concept has the same gondola and mini probe design as the VFA design. The CFA balloon targets a fixed-float altitude of around 52 km. There will naturally be small variations in flight altitude due to atmospheric disturbances. The CFA has a lower balloon envelope mass than the VFA, with a 4 m diameter super-pressure envelope, and requires only 3 inflation tanks. The total estimated aerial platform mass is reduced to approximately 40 kg. The entry probe mass is 120 kg, and the total stacked configuration mass is 520 kg (see the values in Table 3 to compare with the VFA mission). The reduction in mass is not significant compared to the mass margin, but the complexity of balloon operations is reduced considerably.

## 9. Conclusions

We propose a balloon mission to explore the cloud deck of Venus in search of signs of life and to investigate the habitability of cloud particles. The balloon explores the cloud region from 48 to 60 km—completing the baseline mission after two weeks, with the possibility of extended operations lasting up to one month. Science data is returned via an orbiting relay and direct-to-Earth transmission. The balloon deploys miniature probes at select times during its flight operations to provide spatial and temporal diversity in acquired data.

The science payload is a mix of novel high TRL and low TRL instruments. The unique astrobiology-focused goals of the mission, combined with the challenging operational environment, require the capability advancement of instruments that have not yet been applied to planetary exploration. Such instruments, carried by a variable float-altitude balloon, would enable the characterization of habitability and poorly understood phenomena in the Venus clouds and could possibly result in the discovery of signs of life at Venus—which, if confirmed, would be among the most significant events in human history.

**Supplementary Materials:** Not applicable.

**Author Contributions:** Conceptualization, S.S., and S.J.S.; Methodology, and analysis, R.A., W.B., A.A., A.P.G., and M.d.J.; Investigation, R.A., W.B., A.A., A.P.G., M.d.J., S.S., J.J.P., S.J.S., C.E.C., and D.H.G.; writing—original draft preparation, R.A., W.B., J.J.P., S.S., M.d.J., and C.E.C.; writing—review, and editing, J.J.P., S.S., M.d.J., W.B., A.P.G., J.M.L., S.J.S., C.E.C., and D.H.G.; supervision, S.S., J.J.P., and S.J.S.; funding acquisition, S.S. All authors have read and agreed to the published version of the manuscript

**Funding:** This research was partially funded by Breakthrough Initiatives and the Massachusetts Institute of Technology.

**Institutional Review Board Statement:** Not applicable.

**Informed Consent Statement:** Not applicable.

**Data Availability Statement:** Not applicable.

**Acknowledgments:** We thank the extended Venus Life Finder Mission team for useful discussions. List of the individuals involved as the VLF extended Venus Life Finder Mission team can be found here: https://venuscloudlife.com/.

**Conflicts of Interest:** The authors declare no conflict of interest.

## References

1. Greaves, J.S.; Richards, A.M.S.; Bains, W.; Rimmer, P.B.; Sagawa, H.; Clements, D.L.; Seager, S.; Petkowski, J.J.; Sousa-Silva, C.; Ranjan, S.; et al. Phosphine gas in the cloud decks of Venus. *Nat. Astron.* **2020**, *5*, 655–664, https://doi.org/10.1038/s41550-020-1174-4.
2. Bains, W.; Petkowski, J.J.; Seager, S.; Ranjan, S.; Sousa-Silva, C.; Rimmer, P.B.; Zhan, Z.; Greaves, J.S.; Richards, A.M.S. Phosphine on Venus Cannot Be Explained by Conventional Processes. *Astrobiology* **2021**, *21*, 1277–1304. https://doi.org/10.1089/AST.2020.2352/ASSET/IMAGES/LARGE/AST.2020.2352_FIGURE10.JPEG.
3. Morowitz, H.; Sagan, C. Life in the Clouds of Venus? *Nature* **1967**, *215*, 1259–1260. https://doi.org/10.1038/2151259a0.
4. Knollenberg, R.G.; Hunten, D.M. The Microphysics of the Clouds of Venus: Results of the Pioneer Venus Particle Size Spectrometer Experiment. *J. Geophys. Res.* **1980**, *85*, 8039. https://doi.org/10.1029/ja085ia13p08039.
5. Titov, D.V.; Ignatiev, N.I.; McGouldrick, K.; Wilquet, V.; Wilson, C.F. Clouds and Hazes of Venus. *Space Sci. Rev.* **2018**, *214*, 126. https://doi.org/10.1007/s11214-018-0552-z.
6. Oyama, V.I.; Carle, G.C.; Woeller, F.; Pollack, J.B.; Reynolds, R.T.; Craig, R.A. Pioneer Venus Gas Chromatography of the Lower Atmosphere of Venus. *J. Geophys. Res. Earth Surf.* **1980**, *85*, 7891–7902.
7. Mukhin, L.M.; Gel'man, B.G.; Lamonov, N.I.; Mel'nikov, V.V.; Nenarokov, D.F.; Okhotnikov, B.P.; Rotin, V.A.; Khokhlov, V.N. Venera 13 and Venera 14 Gas-Chromatography Analysis of the Venus Atmosphere Composition. *Sov. Astron. Lett.* **1982**, *8*, 399–403.
8. Rimmer, P.B.; Jordan, S.; Constantinou, T.; Woitke, P.; Shorttle, O.; Hobbs, R.; Paschodimas, A. Hydroxide Salts in the Clouds of Venus: Their Effect on the Sulfur Cycle and Cloud Droplet PH. *Planet. Sci. J.* **2021**, *2*, 133. https://doi.org/10.3847/PSJ/ac0156.
9. Mogul, R.; Limaye, S.S.; Way, M.J.; Cordova, J.A. Venus' Mass Spectra Show Signs of Disequilibria in the Middle Clouds. *Geophys. Res. Lett.* **2021**, *48*, e2020GL091327. https://doi.org/10.1029/2020GL091327.

10. Seager, S.; Petkowski, J.J.; Carr, C.E.; Grinspoon, D.; Ehlmann, B.; Saikia, S.J.; Agrawal, R.; Buchanan, W.; Weber, M.U.; French, R.; et al. Venus Life Finder Mission Study. **2021**. https://doi.org/10.48550/arxiv.2112.05153.

11. Sagdeev, R.Z.; Linkin, V.M.; Blamont, J.E.; Preston, R.A. The VEGA Venus Balloon Experiment. *Science* **1986**, *231*, 1407–1408. https://doi.org/10.1126/science.231.4744.1407.

12. Ragent, B. Results of the VEGA 1 Balloon Nephelometer Experiment. *Adv. Sp. Res.* **1987**, *7*, 315–322. https://doi.org/10.1016/0273-117790238-9.

13. Grinspoon, D.; Tahu, G. *Venus Climate Mission Concept Study*; Technical Report, 2010.

14. Wilson, C.F.; Chassefière, E.; Hinglais, E.; Baines, K.H.; Balint, T.S.; Berthelier, J.J.; Blamont, J.; Durry, G.; Ferencz, C.S.; Grimm, R.E.; et al. The 2010 European Venus Explorer (EVE) Mission Proposal. *Exp. Astron.* **2012**, *33*, 305–335. https://doi.org/10.1007/s10686-011-9259-9.

15. Dicicco, A.G.; Nock, K.T.; Powell, G.E. Balloon Experiment at Venus (BEV). In Proceedings of the 11th Lighter-than-Air Systems Technology Conference, Clearwater Beach, FL, USA, 15–18 May 1995; pp. 144–154. https://doi.org/10.2514/6.1995-1633.

16. Sadin, S.R.; Frank, R.E. Buoyant Venus Station Requirements. *J. Spacecr. Rockets* **1970**, *7*, 905–911. https://doi.org/10.2514/3.30069.

17. Beauchamp, P.; Gilmore, M.S.; Lynch, R.J.; Sarli, B.V.; Nicoletti, A.; Jones, A.; Ginyard, A.; Segura, M.E. Venus Flagship Mission Concept: A Decadal Survey Study. In Proceedings of the 2021 IEEE Aerospace Conference, Big Sky, MT, USA, 6–13 March 2021. https://doi.org/10.1109/AERO50100.2021.9438335.

18. Bains, W.; Petkowski, J.J.; Rimmer, P.B.; Seager, S. Production of Ammonia Makes Venusian Clouds Habitable and Explains Observed Cloud-Level Chemical Anomalies. *Proc. Natl. Acad. Sci. USA.* **2021**, *118*, e2110889118. https://doi.org/10.1073/pnas.2110889118.

19. Limaye, S.S.; Mogul, R.; Smith, D.J.; Ansari, A.H.; Słowik, G.P.; Vaishampayan, P. Venus' Spectral Signatures and the Potential for Life in the Clouds. *Astrobiology* **2018**, *18*, 1181–1198. https://doi.org/10.1089/ast.2017.1783.

20. Wu, Z.; Wan, H.; Xu, J.; Lu, B.; Lu, Y.; Eckhardt, A.K.; Schreiner, P.R.; Xie, C.; Guo, H.; Zeng, X. The Near-UV Absorber OSSO and Its Isomers. *Chem. Commun.* **2018**, *54*, 4517–4520. https://doi.org/10.1039/C8CC00999F.

21. Kovtunenko, V.M.; Prospekt, L.; Sagdeev, R.Z.; Barsukov, V.L. VEGA Project Re-Entry Vehicle of "VEGA" Spacecraft. *Acta Astronaut.* **1986**, *13*, 425–432.

22. Asmar, S.; Matousek, S. Mars Cube One (MarCO) Shifting the Paradigm in Relay Deep Space Operations. In Proceedings of the 14th International Conference on Space Operations, Daejeon, Korea, 16–20 May 2016. https://doi.org/10.2514/6.2016-2483.

23. Bienstock, B.J. Pioneer Venus and Galileo Entry Probe Heritage. In Proceedings of the Planetary Probe Atmospheric Entry and Descent Trajectory Analysis and Science, Lisbon, Portugal, 7 October 2003.

24. Rodier, R.; Thuss, R.; Terhune, J. Parachute Design for Galileo Jupiter Entry Probe. In Proceedings of the 7th Aerodynamic Decelerator and Balloon Technology Conference, San Diego, CA, USA, 21–23 October 1981; pp. 1951. https://doi.org/10.2514/6.1981-1951.

25. EP1886920A1–Flexible Vessel–Google Patents. Available online: https://patents.google.com/patent/EP1886920A1 (accessed 2022-03-06).

26. De Jong, M. Venus Altitude Cycling Balloon. *Venus Sci. Priorities Lab. Meas.* **2015**, *1838*, 4030.

27. Fimmel, R.O.; Colin, L.; Burgess, E. *Pioneer Venus*; National Aeronautics and Space Administration: Washington, DC, USA, 1983.

28. Balint, T.S.; Kolawa, E.A.; Cutts, J.A.; Peterson, C.E. Extreme Environment Technologies for NASA's Robotic Planetary Exploration. *Acta Astronaut.* **2008**, *63*, 285–298. https://doi.org/10.1016/j.actaastro.2007.12.009.